\documentstyle[aps,prl,twocolumn]{revtex}
\input{epsf}
\begin{document}
\draft
\title{Search for correlation effects in linear chains of trapped ions}
\author{C. J. S. Donald, D. M. Lucas, P. A. Barton$^{*}$, M. J. McDonnell, J. P. Stacey, D. A. Stevens$^{\dag}$, D. N. Stacey, and A.~M.~Steane \\
\small$^{*}$Institut f\"{u}r Experimentalphysik, Universit\"{a}t Innsbruck,
Technikerstr. 25, A-6020 Innsbruck, Austria
\\ $^{\dag}$L'Institut d'Optique, Universite de Paris Sud, Orsay, France  \\
 Centre for Quantum Computation, Department of Atomic and Laser Physics,
University of Oxford,\\ \small Clarendon Laboratory, Parks Road, Oxford OX1 3PU, U.K.}
\date{\today}
\maketitle
\begin{abstract}
We report a precise search for correlation effects in linear chains of 2 and 3 trapped
Ca$^{+}$ ions.  Unexplained correlations in photon emission times within a linear
chain of trapped ions have been reported, which, if genuine, cast doubt on the
potential of an ion trap to realize quantum information processing.  We observe
quantum jumps from the metastable 3d$^{2}$D$_{5/2}$ level for several hours, searching
for correlations between the decay times of the different ions.  We find no evidence
for correlations: the number of quantum jumps with separations of less than 10~ms is
consistent with statistics to within errors of $0.05\%$; the lifetime of the
metastable level derived from the data is consistent with that derived from
independent single-ion data at the level of the experimental errors ($1\%$); and no
rank correlations between the decay times were found with sensitivity to rank
correlation coefficients at the level of $|R| = 0.024$.
\end{abstract}
\pacs{ 42.50.Lc, 42.50.Fx, 32.80.Pj, 32.70.Cs}
%
%
The drive to realise the potential of quantum information processing
\cite{98:roysoc,98:steane} has led to the investigation of various experimental
systems; among these is the ion trap, which has several advantages including the
capability to generate entanglement actively with existing technology
\cite{00:sackett}.  Following the proposal of an ion-trap quantum processor by Cirac
and Zoller \cite{95:cirac}, several groups have carried out pioneering experiments
\cite{95:monroe,98:king,99:nagerl,99:roos,99:peik}. In a recent review
\cite{98:hughes}, the view was expressed that ``the ion trap proposal for realizing a
practical quantum computer offers the best chance of long term success.'' One of the
attractive features of the trap is that the various interactions and processes which
govern its behaviour have been exhaustively studied and are in principle
well-understood.  However, 14 years ago unexplained collective behaviour when several
ions were present was reported \cite{86:sauter}. This prompted tests in another
laboratory which gave null results \cite{88:itanoA,88:itanoB}, but recently a further
account of such effects has appeared \cite{99:block}.   There is thus an apparent
conflict of evidence from different laboratories.

The effects manifest themselves as an enhanced rate of coincident quantum jumps.
Sauter $et$ $al.$ \cite{86:sauter} measured two- and three-fold coincident quantum
jumps in a system of three trapped Ba$^{+}$ ions to occur two orders of magnitude more
frequently than expected on the basis of statistics. This observation led to proposals
that the ions were undergoing a collective interaction with the light field
\cite{86:sauter,88:Blatt}. Itano {\em et al.} \cite{88:itanoA,88:itanoB} subsequently
made a search for such effects in groups of two and three Hg$^{+}$ ions in their
laboratory. Their results were consistent with no correlations. In a test on two ions,
when over 5649 consecutive jumps were observed, the number of apparent double jumps
was 11, which was approximately the number that would be expected due to random
coincidences within the finite time resolution of the experiment. Further tests based
on photon statistics were also consistent with no correlations.

More recently, Block $et$ $al.$ \cite{99:block} have observed an enhanced rate of two-
and three-fold coincidences in a linear chain of ten Ca$^{+}$ ions, where the
coincidences were not confined to adjacent ions.  This led them to suggest an
unexplained long range interaction between ions in the linear crystal.  They also
found that measurements of the lifetime $\tau$ of the 3D$_{5/2}$ level (shelved state)
from the 10-ion string produced discrepancies of as much as $6\sigma$ between runs
under nominally identical conditions, where $\sigma$ is the standard deviation for
each run.

Since only the electromagnetic interaction is involved, it is extremely unlikely that
these observations indicate new physics; nevertheless, they raise serious doubt about
the suitability of the ion trap as a quantum information processing device. The
coupling between a quantum system and its environment plays a crucial role in quantum
information processing. An unexplained contribution to this coupling is especially
significant, because any method to suppress the decoherence, such as quantum error
correction (QEC) \cite{98:steane}, relies on accurate knowledge of the process in
question. It is furthermore particularly important to understand collective
decoherence processes and place a reliable upper bound on their size, because the
simultaneous combination of uncorrelated and correlated errors in a quantum computer
poses the most severe constraints on QEC \cite{77:macwilliams}. Thus, experimental
reports of a decoherence process which is both unexplained and collective merit
serious attention. We have therefore undertaken a search for the reported effects in
linear chains of 2 and 3 trapped Ca$^+$ ions.

Our data were taken under conditions such that correlation effects would be expected
on the basis of the results of \cite{86:sauter} and \cite{99:block}, and are
significantly more precise than either. We find no evidence at all for correlations.
Our work is complementary to that of \cite{88:itanoA,88:itanoB} in that we are
operating in a different system (Ca$^+$ instead of Hg$^+$) with a significantly
different time-scale (mean rate for observed double quantum jumps of order 0.2 per
minute instead of 2 per minute), and we perform several new statistical tests on 2 and
3 ions. Our upper bound for unexpected double jumps is $1.4$ per hour, or $0.05$\% of
the single jump rate. The corresponding upper bounds for the Hg$^+$ ion trap in
\cite{88:itanoB} are 30 per hour and $0.06$\%.

\begin{figure}[tbp]
\begin{center} \mbox{ \epsfxsize 8.5cm\epsfbox{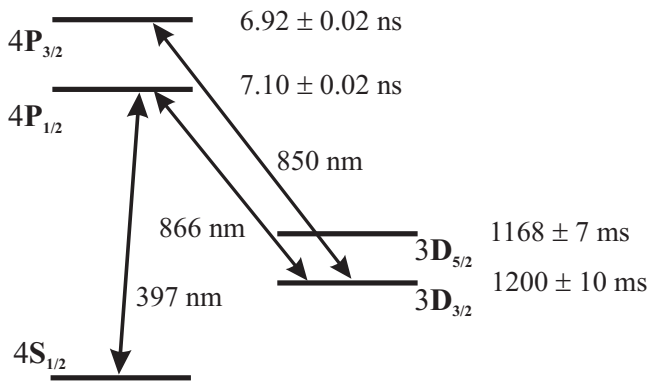}} \end{center}
\caption{Low-lying energy levels of $^{40}\mbox{Ca}^{+}$, with their lifetimes. Lasers
at 397~nm, 866~nm and 850~nm drive the corresponding transitions in the
experiments.}\label{fig:levels}
\end{figure}

The experimental method is very similar to that reported in our measurement of the
lifetime of the 3d$^{2}$D$_{5/2}$ level \cite{00:barton}, which was originally adopted
by Block $et$ $al.$ \cite{99:block}.  Linear crystals of a small number, $N$, of
$^{40}$Ca$^{+}$ ions separated by about 15~$\mu$m are obtained by trapping in a linear
Paul trap $in$ $vacuo$ ($\leq 2\times 10^{-11}$~Torr), and laser-cooling the ions to a
few mK. The transitions of interest are shown in figure \ref{fig:levels}. Laser beams
at 397 nm and 866 nm continuously illuminate the ions, and the fluorescence at 397 nm
is detected by a photomultiplier. The photon count signal is accumulated for bins of
duration $t_{b} = 10.01$~ms (of which the last 2.002 ms is dead time), and logged. A
laser at 850 nm drives the $3\mbox{D}_{3/2} - 4\mbox{P}_{3/2}$ transition. The most
probable decay route from $4\mbox{P}_{3/2}$ is to the $4\mbox{S}_{1/2}$ ground state;
alternatively, an ion can return to $3\mbox{D}_{3/2}$. However, about 1 decay in 18
occurs to $3\mbox{D}_{5/2}$, the metastable ``shelving'' level. At this point the
fluorescence from the ion that has been shelved disappears.  A shutter on the 850 nm
laser beam remains open for 100 ms before it is closed, which gives ample time for
shelving of all $N$ ions. Between 5 and 10 ms after the shutter is closed we start to
record the photomultiplier count signal in the 10 ms bins. We keep observing the
photon count until it abruptly increases to a level above a threshold. This is set
between the levels observed when 1 and 0 ions remain shelved. The signature for all
$N$ ions having decayed is taken to be ten consecutive bins above this threshold.
After this we re-open the shutter on the 850 nm laser.  This process is repeated for
several hours, which constitutes one run.

\begin{figure}[tbp]
\begin{center} \mbox{ \epsfxsize 8.5cm\epsfbox{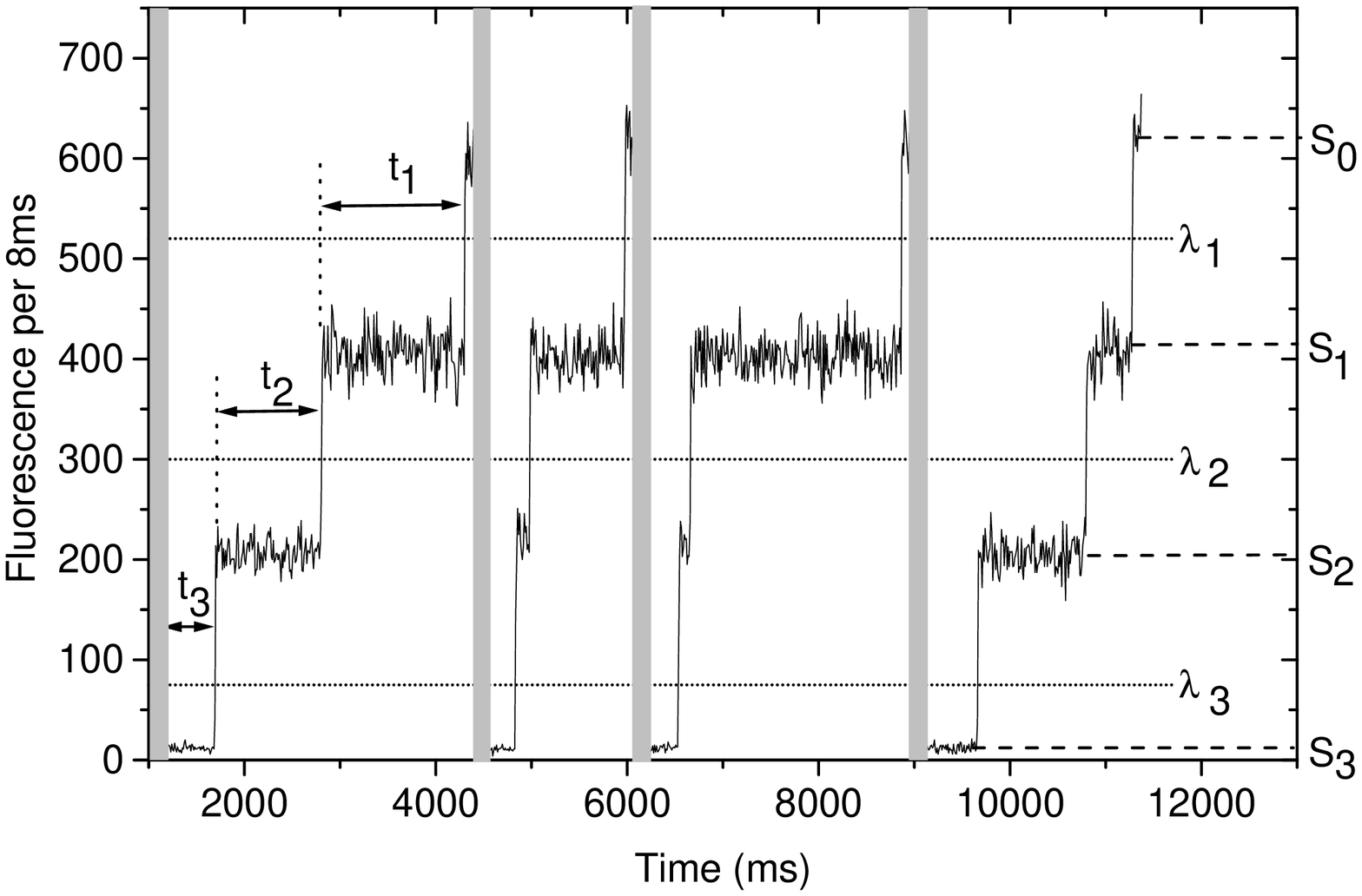}} \end{center}
\caption{Observed fluorescence signals from a linear 3-ion crystal.  The vertical axis
is the number of counts given by the photomultiplier during one 10 ms counting bin (2
ms dead time).  The grey bars indicate re-shelving periods, when the shutter on the
850 nm laser was open. The de-shelving times, $t_{m}$, are labelled for one
observation of the 3 ions decaying from the shelved state, where $m$ is the number of
ions remaining to decay. The dotted horizontal lines show the threshold settings
$\lambda_{m}$ for the data analysis; the dashed horizontal lines show the mean count
levels $S_{m}$.}
 \label{fig:raw}
\end{figure}

\begin{figure}[tbp]
\begin{center} \mbox{ \epsfxsize 8.5cm\epsfbox{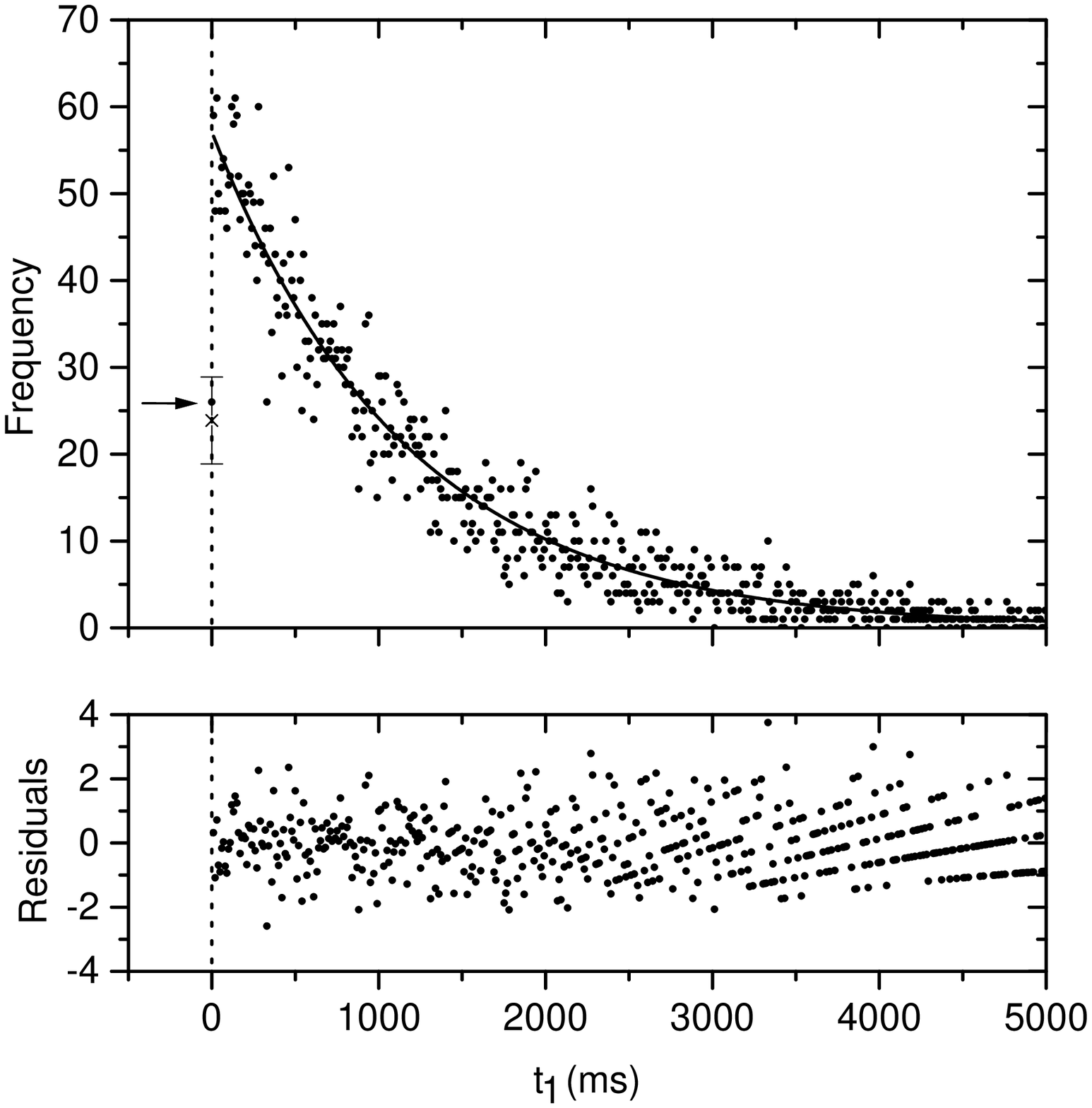}} \end{center}
\caption{The histogram of the decay times, $t_{1}$, of the last ion of 2 to decay,
obtained from a 3.2 hour run, with an exponential $A \exp{(-\gamma_1 t)}$ fitted to
all bins but the first two. In this case, the analysis gave $ A = 57 \pm 1$,
$\gamma_{1} = 0.860 \pm 0.012$~s$^{-1}$, which agrees with the expected rate
$\gamma_{1} = 1/\tau = 0.856\pm 0.005$~s$^{-1}$, where $\tau$ is the lifetime derived
from single-ion data \protect\cite{00:barton}.  The residuals are shown on an expanded
scale, in the form (data$-$fit)/$\sqrt{\mbox{fit}}$.  The first bin gives the number
of 2-ion jumps observed to be coincident within one counting bin and has a modified
bin width (see text), which reduces the expected number in the first bin to be
$F=0.42$ of the value, $f_{1} = 57$, predicted by the fitted exponential. The expected
number, $F f_{1} = 24 \pm 5$ (marked with a cross), agrees with the observed number,
$26$ (indicated by an arrow).}
 \label{fig:hist}
\end{figure}

The data from a given run were analysed as follows.  The raw data consists of counts
indicating the average fluorescence level in each bin of duration $t_{b}$  (see figure
\ref{fig:raw}).  $N$ thresholds $\lambda_{m}$ are set, the $m^{th}$ threshold being
set between the levels observed when $m$ and $(m-1)$ ions remain shelved. The number
of bins observed below $\lambda_{N}$ gives the decay time, $t_{N}$, of the first of
$N$ shelved ions to decay.  The number of bins observed between $\lambda_{m+1}$ and
$\lambda_{m}$ being exceeded gives the decay time, $t_{m}$, of the next ion to decay
leaving $(m-1)$ ions shelved.  The large number of $t_{m}$ obtained are then gathered
into separate histograms and the expected exponential distribution $A \exp{(-\gamma_m
t)}$ is fitted to each, in order to derive the decay rate $\gamma_{m}$ of the next ion
to decay leaving $(m-1)$ ions shelved (see figure \ref{fig:hist}).  It is appropriate
to use a Poissonian fitting method (described in \cite{00:barton}), rather than
least-squares, because of the small numbers involved in part of the distribution (at
large $t$).


If the $N$ ions are acting independently, each one will have a decay rate $\gamma =
1/\tau$, where $\tau$ is the lifetime of the 3D$_{5/2}$ state.  Since we do not
distinguish between the fluorescence signals from the different ions, then with $m$
ions remaining shelved the next decay is characterised by the increased rate
$\gamma_{m} = m/\tau$.

Figure \ref{fig:hist} shows the histogram of the decay times, $t_{1}$, of the second
ion of two to decay obtained from a 3.2 hour run.  The expected exponential decay fits
the data very well. Events in the first bin of the histogram correspond to both ions
being detected as decaying in the same bin, $t_{1} = 0$. These quantum jumps,
coincident within our time resolution, certainly do not occur two orders of magnitude
more frequently than expected by random coincidence as was observed by Sauter $et$
$al.$ \cite{86:sauter}. In fact, they are observed to occur less frequently than
predicted by the fitted exponential to the histogram data. However, this is an
artefact of our finite time resolution.  The fitted exponential to the histogram data
has value $f_{1}$ in the first bin, which gives the number of second ion decays that
are expected to occur within $t_{b}$ of the first ion decaying by random coincidence.
However, for both ions to decay within a single bin, the second ion has an average
time of less than $t_{b}$ in which to decay. The exact details depend upon the
analysis thresholds, $\lambda_{m}$, and the detector dead time. In the 2-ion case, one
can show that, to first order in $t_b / \tau$, the first bin width is modified to $F
t_{b}$ where:
\[
F = 0.98 - 0.8\lambda'_1 + 0.16{\lambda'_1}^2 + 0.16{\lambda'_2}^2 + 1.44\lambda'_2 -
0.64\lambda'_1\lambda'_2
\]
with normalized thresholds:
\[ \lambda'_m = \frac{\lambda_m - S_N}{S_{N-1} - S_N} \]
where $S_{m}$ is the mean photon count with $m$ ions shelved (so $S_N$ is the mean
background count level).  This expression was verified using real and simulated data.
The expected number of coincidences is therefore $F f_{1}$. For the histogram shown,
the 2-ion data was analyzed with  the thresholds $\lambda'_1 = 1.4$ and $\lambda'_2 =
0.40$ (these are chosen to optimize the discrimination of the fluorescence levels
$S_{m}$), which gives $F=0.42$. The expected number of coincidences is $F f_{1} = 24
\pm 5$, assuming $\sqrt{n}$ errors, which agrees with the observed number of
coincidences, 26.  The second bin of the histogram is the only other bin expected to
have a modified width, which is by a negligible amount.  Note that, to ensure the
number of coincidences is properly normalized, it is important that only events where
at least $(m+1)$ ions were shelved at the start of an observation are included in the
$t_m$ histogram (for $m\ne N$).

Table \ref{tab:coinc} shows that the observed number of 2-fold coincidences in the 2-
and 3-ion data agree with the expected value within $\sqrt{n}$ errors. The total
expected number of 2-fold coincidences in all the data was 66.3 out of 16132 quantum
jumps observed to start with at least 2 ions shelved.  We are therefore sensitive to
changes in the proportion of 2-fold coincidences at the level of $\sqrt{66}/16132 =
0.05\%$ or about 1.4 event per hour.

The expected number of 3-fold coincidences depends on the threshold settings in a more
complex way than in the 2-fold case, and here we simply use simulated 3-ion data to
provide the predicted number of 3-fold coincidences shown in table \ref{tab:coinc}.
The total number of expected 3-fold coincidences is 0.05 in both 3-ion data runs,
which have a combined duration of 2.8 hrs. In fact, this predicted value is
significantly lower than effects in our trap which can perturb the system sufficiently
to cause de-shelving (such as collisions with residual background gas), as discussed
in \cite{00:barton}.  We observe at most one event, depending on the exact choice of
threshold settings, and this does not constitute evidence for correlation.

\begin{figure}[tbp]
\begin{center} \mbox{ \epsfxsize 8.5cm\epsfbox{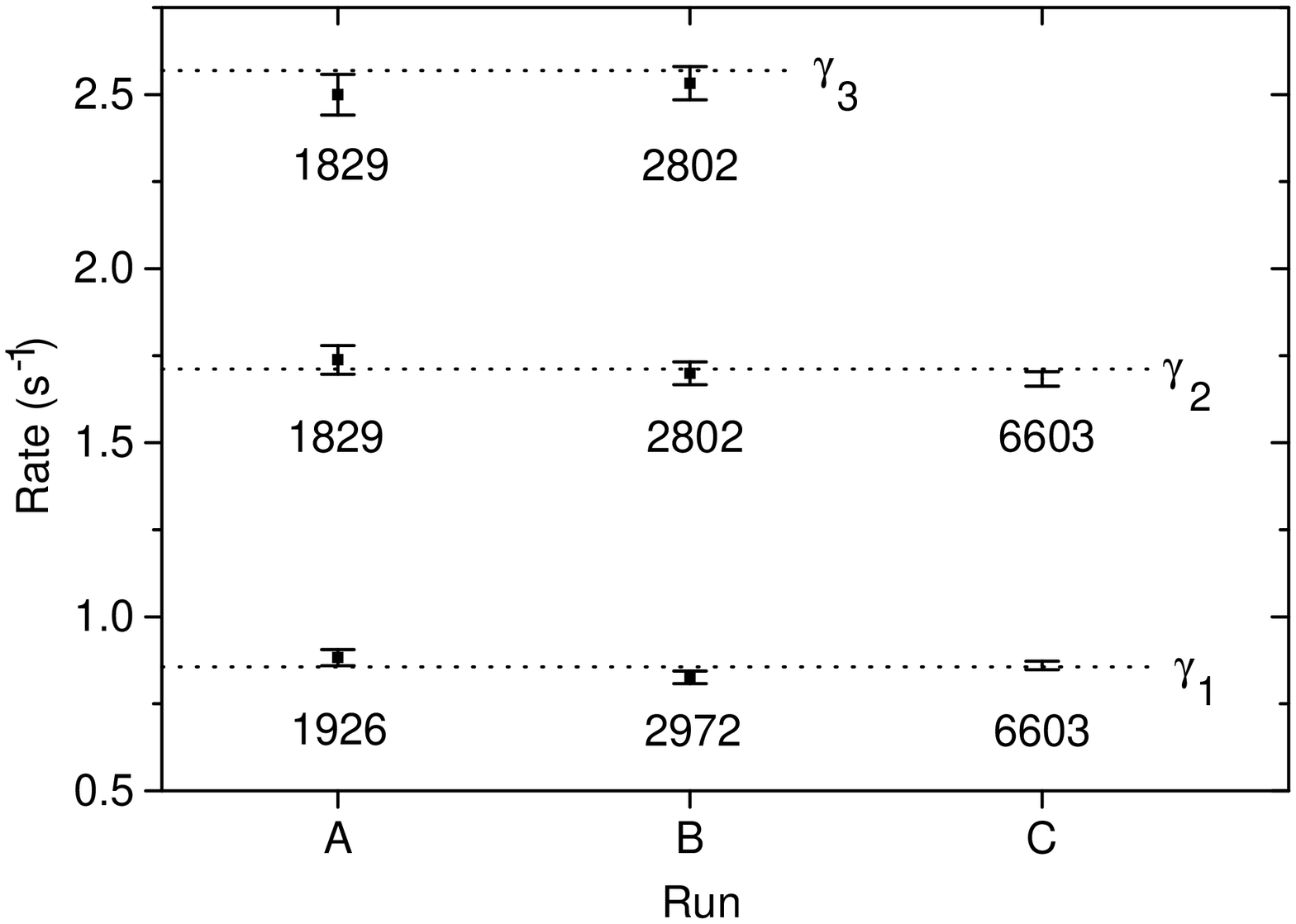}} \end{center}
\caption{ Measured de-shelving rates $\gamma_{m}$ of the the next ion to decay from
the state where $m$ ions are shelved; errors are purely statistical. The horizontal
lines are the expected rates $\gamma_{m} = m/\tau$ if the ions are acting
independently, where $\tau$ is the lifetime derived from single-ion data
\protect\cite{00:barton} and have negligible error on this scale. Runs A and B were
conducted with 3 ions, run C with 2 ions.  The number below each point gives the
number of decay times in the corresponding histogram.  }
 \label{fig:rates}
\end{figure}

The decay rates obtained from the 2- and 3-ion data are shown in figure
\ref{fig:rates}, where the horizontal lines are the expected rates $\gamma_{m} =
m/\tau$ assuming the ions to act independently. Combining all the $\gamma_{m}$ derived
from the 2- and 3-ion data as estimates of $m/\tau$ yields a value $\tau = 1177 \pm
10$~ms, where we include a 2~ms allowance for systematic error \cite{00:barton}. This
is consistent with the value derived from single-ion data, $\tau = 1168 \pm 7$~ms
\cite{00:barton}.  We are therefore sensitive to changes in the apparent value of
$\tau$ due to multiple ion effects at the level of 1\%.  Superfluorescence and
subfluorescence as observed in a two-ion crystal \cite{96:devoe} are calculated to be
negligible with the large interionic distance of about 15~$\mu$m in the chain.

In order to look for more general forms of correlation between the decay times of each
ion, rank correlation tests were performed.  Table~\ref{tab:cor} gives the results;
they show no significant correlations.  The 2-ion data is the most sensitive, allowing
underlying rank-correlation coefficients to be ruled out at the level of $|R_{12}| =
0.024$.


In summary, we have presented results that are consistent with no correlations of
spontaneous decay within linear chains of 2 and 3 trapped Ca$^{+}$ ions, contrary to
previous studies. First, the number of coincident quantum jumps were found to be
consistent with those expected from random coincidence at the level of $0.05\%$.
Second, the exponential decay expected assuming the ions to act independently fitted
the histogram of decay times $t_{m}$ obtained from the 2- and 3-ion data well.  Third,
the decay rates from these fits were combined to estimate the lifetime of the shelved
state, giving a result consistent with our previous precise measurement performed on a
single ion \cite{00:barton}. Fourth, rank correlation tests were performed on the
decay times obtained from the 2- and 3-ion data; no evidence for rank correlation was
found.

We suggest therefore that the correlations which have been reported are likely to be
due not to interactions between the ions themselves, but to external time-dependent
perturbations.  In our own trap, we have investigated and reduced such perturbations
to a negligible level \cite{00:barton}, and the present work demonstrates that when
this is done there is no evidence that an ion trap is subject to unexplained effects
which would make it unsuitable for quantum information processing.

We are grateful to G.R.K.~Quelch for technical assistance, and to S.~Siller for useful
discussions. This work was supported by EPSRC (GR/L95373), the Royal Society, Oxford
University (B(RE) 9772) and Christ Church, Oxford.
%

%
\begin{table}[tbp]
\begin{center}
\begin{tabular}{ddd@{\hspace{0em}}dddd}
Run & $N$ & Time (hrs) & $m_{i}\rightarrow m_{f}$ & $N_{\mbox{\tiny{QJ}}}$ &
$n_{\mbox{\tiny{c}}}$ & $n_{\mbox{\tiny{obs}}}$ \\ \hline \\[-1ex] %
  &   &     & $2\rightarrow 0$ & 1926 & 7.0 & 10 \\ %
A & 3 & 1.1 & $3\rightarrow 1$ & 1829 & 9.5 & 9 \\
  &   &     & {\bf 3~$\rightarrow$~0} & {\bf 1829} & {\bf 0}.{\bf 02} & {\bf 0} \\
  &   &     &       &      &      &  \\[-1ex]
  &   &     & $2\rightarrow 0$ & 2972 & 10.5 & 13 \\ %
B & 3 & 1.7 & $3\rightarrow 1$ & 2802 & 15.4 & 13 \\
  &   &     & {\bf 3~$\rightarrow$~0} & {\bf 2802} & {\bf 0}.{\bf 03} & {\bf 0} \\
  &   &     &       &      &      &  \\[-1ex]
C & 2 & 3.2 & $2\rightarrow 0$ & 6603 & 23.9 & 26 \\
  &   &     &       &      &      &  \\[-1ex] \hline
\multicolumn{2}{l}{total 2-fold} & 6.0 & $(2,3)\rightarrow (0,1)$ & 16132 & 66.3 & 71
\\
\multicolumn{2}{l}{\bf total 3-fold} & {\bf 2}.{\bf 8} & {\bf 3~$\rightarrow$~0} &
{\bf 4631} & {\bf 0}.{\bf 05} & {\bf 0}
\\
\end{tabular}
\end{center}
\caption{Two-fold and three-fold (bold type) coincident quantum jumps, with $N$ ions.
Coincident quantum jumps occur with $m_{i}$ ions initially shelved, leaving $m_{f}$
ions shelved.  $N_{\mbox{\tiny{QJ}}}$ is the total number of quantum jumps observed
with $m_{i}$ ions initially shelved.  For independent ions, $n_{\mbox{\tiny{c}}}$ of
these jumps are predicted to be coincident, taking into account the modified bin
width. $n_{\mbox{\tiny{obs}}}$ gives the number of coincidences observed.  The third
column gives the total amount of time that one or more ions spent shelved in each
run.}
 \label{tab:coinc}
\end{table}
\begin{table}[tbp]
\begin{center}
\begin{tabular}{dddddc}
Run & $N$ & $R_{12}$ & $R_{23}$ & $R_{13}$ & $R^{95\%}$
\\ \hline
A & 3 & $-$0.025 & $-$0.010 & $-$0.018 & 0.046 \\ %
B & 3 & $-$0.019 & $+$0.010 & $+$0.008 & 0.037 \\%
C & 2 & $+$0.008 &    ---   &    ---   & 0.024 \\
\end{tabular}
\end{center}
\caption{ Results of the rank correlation tests, with $N$ ions.  $R_{nm}$ is the
Spearman rank-order correlation coefficient for the decay times $t_{n}$ and $t_{m}$.
$|R_{nm}|$ would have to be greater than $R^{95\%}$ for 95\% significance
\protect\cite{numrec}.}
 \label{tab:cor}
\end{table}
\end{document}